# Linear mixed models to handle missing at random data in trial-based economic evaluations

Andrea Gabrio[1], Catrin Plumpton[2], Sube Banerjee[3], Baptiste Leurent[4]

**Abstract**: Trial-based cost-effectiveness analyses (CEAs) are an important source of evidence in the assessment of health interventions. In these studies, cost and effectiveness outcomes are commonly measured at multiple time points, but some observations may be missing. Restricting the analysis to the participants with complete data can lead to biased and inefficient estimates. Methods, such as multiple imputation, have been recommended as they make better use of the data available and are valid under less restrictive Missing At Random (MAR) assumption. Linear mixed effects models (LMMs) offer a simple alternative to handle missing data under MAR without requiring imputations, and have not been very well explored in the CEA context. In this manuscript, we aim to familiarise readers with LMMs and demonstrate their implementation in CEA. We illustrate the approach on a randomised trial of antidepressant, and provide the implementation code in R and Stata. We hope that the more familiar statistical framework associated with LMMs, compared to other missing data approaches, will encourage their implementation and move practitioners away from inadequate methods.

## 1 INTRODUCTION

Cost-effectiveness analyses (CEAs) conducted alongside randomised controlled trials are an important source of information for decision-makers in the process of technology appraisal (Ramsey et al., 2015). The analysis is based on healthcare outcome data and health service use, typically collected at multiple time points and then combined into overall measures of effectiveness and cost (Drummond et al., 2005). The derivation of these measures is problematic in the presence of missing outcome data, which are a common issue in trials, with on average only 60% to 75% of randomised participants having complete CEA data (Leurent et al., 2020; Noble et al., 2012). A popular approach to handle missingness is to discard the participants with incomplete observations (*complete case analysis* or CCA), allowing for derivation of the overall measures based on the completers alone. We note that slightly different definitions of CCA are possible , depending on the form of the model of interest, the type of missingness and the inclusion of observed covariates. Without loss of generality, throughout the paper, we refer to CCA as to when only individuals with fully-observed effect and cost data (i.e. the completers) are included in the analysis, thus requiring the removal of all cases with partially-observed outcome data. This approach, although appealing by its simplicity, has well-recognised limitations including loss of efficiency and an increased risk of bias (Carpenter & Kenward, 2012; Faria et al., 2014; Rubin, 1987). In recent years there has been an increment in the uptake of more appropriate statistical methods, such as multiple imputation (Leurent, Gomes, Faria, et al., 2018) or Bayesian methods (Gabrio et al., 2019), which improve efficiency, rely on less restrictive missingness assumptions, and facilitate the task of conducting sensitivity analysis to alternative missingness assumptions ((Gabrio et al., 2020; Leurent et al., 2020). Among these, Missing At Random [MAR; Rubin (1987)] often provides a desirable starting point as it implies that valid inferences can be drawn based on the observed data. Although both CCA and MI are broadly valid under MAR assumptions, the condition for validity of CCA is more restrictive as it requires that individuals with partially-observed data are not systematically different from the completers, which is instead relaxed in MI. Although these methods have become more frequent and accessible in CEAs, their implementation in routine analyses remains limited compared to CCA (Gabrio et al., 2017; Leurent, Gomes, & Carpenter, 2018), perhaps because of a lack of familiarity, computational time, or the analytical difficulties of the approaches. Indeed, these methods require particular care when combined with other statistical procedures, such as taking into account clustering (R. Gomes et al., 2012), or using bootstrap re-sampling (Brand et al., 2019).

---

[1] *Department of Methodology and Statistics, Faculty of Health Medicine and Life Science, Maastricht University, NL*

[2] *Centre for Health Economics and Medicines Evaluation, Bangor University, Bangor, UK*

[3] *Centre for Dementia Studies, Brighton and Sussex Medical School, Brighton, UK*

[4] *Department of Medical Statistics, London School of Hygiene and Tropical Medicine, UK*

We propose the use of *linear mixed effects models* (LMMs) as an alternative approach under MAR. LMMs are commonly used for the modelling of dependent data (e.g. repeated-measures) and belong to the general class of likelihood-based methods. LMMs have been occasionally used in CEA under the denomination of *multilevel* or *hierarchical models* (Manca et al., 2005; Rice & Jones, 1997) to account for the dependence between observations with a hierarchical structure (e.g. cluster-randomised trials). Repeated-measures also follow a hierarchical structure since data within each individual are correlated, and LMM are increasingly used to analyse longitudinal outcome in trials. However, LMMs appear surprisingly uncommon for the analysis of repeated measures in trial-based CEA, perhaps because of a lack of awareness or familiarity with fitting LMMs. To our knowledge, only (Faria et al., 2014) briefly examined the use of LMMs in trial-based CEA as an alternative to multiple imputation, and concluded that they offer a valid approach under MAR.

In this letter, we aim to familiarise readers with the implementation of LMMs in trial-based CEA using standard software and summarise the statistical and economic results from a case study. Finally, we discuss the proposed approach and provide some suggestions for future work.

## 2 METHODS

### 2.1 Linear mixed effects models for repeated measurements

Linear mixed model extends the usual linear model framework by the addition of "random effect" terms, which can take into account the dependence between observations. A simple model for a repeated measure can be written as

$$Y_{ij} = \beta_1 + \beta_2 X_{i1} + \cdots + \beta_{P+1} X_{iP} + \omega_i + \epsilon_{ij}, \quad (1)$$

where $Y_{ij}$ denotes the outcome repeatedly collected for each individual $i = 1, \ldots, N$ at multiple times $j = 1, \ldots, J$. The model parameters commonly referred to as *fixed effects* include an intercept $\beta_1$ and the coefficients $(\beta_2, \ldots, \beta_{P+1})$ associated with the predictors $X_{i1}, \ldots, X_{iP}$, while $\omega_i$ and $\epsilon_{ij}$ are two random terms: $\epsilon_{ij}$ is the usual error term and $\omega_i$ is a *random intercept* which captures variation in outcomes between individuals. The random terms are typically assumed to be normally distributed such that $\omega_i \sim N(0, \sigma_\omega^2)$ and $\epsilon_{ij} \sim N(0, \sigma_\epsilon^2)$. Equation 1 treats the data as having a 2-level structure, where $\sigma_\omega^2$ and $\sigma_\epsilon^2$ capture the variance of the responses within (level 1) and between (level 2) individuals, respectively. The models can be extended to deal with more complex structures, for example by allowing the effect of the covariates to vary across individuals (random slope) or a different covariance structure of the errors. LMMs can be fitted even if some outcome data are missing and provide correct inferences under MAR (Schafer & Graham, 2002). We refer interested readers to the available LMM literature for an in-depth description of the methods (Molenberghs et al., 2004).

A particular type of LMMs commonly used in the analysis of repeated measures in clinical trials is referred to as *Mixed Model for Repeated Measurement* (MMRM) (Wolfinger, 1993). The model includes a categorical effect for time, an interaction between time and treatment arm, and allows errors to have different variance and correlation over time (i.e. unstructured covariance structure). Consider the case where we want to model health related quality of life data (i.e. utilities) collected at three times (baseline and two follow-ups). The model can then be expressed as:

$$U_{i1} = \beta_1 \text{TIME}_1 + \omega_i + \epsilon_{i1}$$

$$U_{i2} = \beta_2 \text{TIME}_2 + \beta_3 \text{TIME}_2 \text{TRT}_i + \omega_i + \epsilon_{i2}, \quad (2)$$

$$U_{i3} = \beta_4 \text{TIME}_3 + \beta_5 \text{TIME}_3 \text{TRT}_i + \omega_i + \epsilon_{i3}$$

where $U_{ij}$ is the utility measured for patient $i$ at time $j$, $\text{TIME}_j$ and $\text{TRT}_i$ are the indicators for the time and treatment arm, and $\text{TIME}_j \text{TRT}_i$ are the interaction terms between time at each follow-up ($j \geq 2$). Within this model, $\beta_1$, represents the mean utility at baseline, $\beta_2$ and $\beta_4$ represent the mean utility for the control arm ($\text{TRT}_i = 0$) at the first and second follow-up, while $\beta_3$ and $\beta_5$ capture the mean difference between the arms at the two follow-ups. The errors at each time follow a multivariate normal distribution $\epsilon_i \sim N(\mathbf{0}, \mathbf{\Sigma})$ with an unstructured covariance matrix $\mathbf{\Sigma}$, i.e. errors are allowed to have different variance and correlation between them. Note that no treatment parameter was included at baseline, this is sometime referred to as an "constrained" model, allowing to obtain parameters adjusted for baseline outcome values (see Section 2.3). The model can be easily extended to handle additional time-points, or to model any other repeatedly measured continuous outcome such as cost.

Incremental (between-group) or marginal (within-group) estimates for aggregated outcomes over the trial period, such as quality-adjusted life years (QALYs) or total costs can be retrieved as linear combinations of the parameter estimates from Equation 2. For example, the mean difference in total cost is obtained by summing up the estimated differences at each follow-up point, while differences on a QALY scale can be obtained as weighted linear combinations of the coefficient estimates of the utility model (see Appendix C).

## 2.2 Adjusting for baseline variables

It is standard practice to adjust the analysis for imbalances in some baseline characteristics, to control for potential imbalance and as well as to increase precision (Manca et al., 2005) while also strengthening the plausibility of the MAR assumption (Little et al., 2012). We note that, differences in baseline outcome values between arms are already adjusted for within the model specification in Equation 2. Other baseline variables can be adjusted for by adding them as predictors to the model, although LMMs require covariates to be completely observed. However, in randomised controlled trials, missing baseline data can be usually addressed by implementing single imputation techniques (e.g. mean-imputation) to obtain complete data prior to fitting the model, without loss of validity or efficiency (White & Thompson, 2005).

## 2.3 Assessing cost-effectiveness

Once estimates for the average effectiveness and total cost differences are obtained, uncertainty can be assessed through resampling methods (e.g. bootstrapping). Results are then often summarised using the *Cost-Effectiveness Plane* [CEP; Black (1990)] and the *Cost-Effectiveness Acceptability Curve* [CEAC; Van Hout et al. (1994)].

# 3 TRIAL OVERVIEW

## 3.1 Overview

The Health Technology Assessment Study of the Use of Antidepressants for Depression in Dementia (HTA-SADD) was a placebo-controlled randomised trial of participants from old-age psychiatry services in England (Banerjee et al., 2011). A total of $n = 326$ participants were enrolled and randomised to receive placebo, or either one of two antidepressants (sertraline and mirtazapine) together with treatment as usual, and followed for 9 months. Outcomes were collected at baseline, 3 months and 9 months, including EQ-5D-3L which was converted into utility (Dolan, 1997), and resource use for $0 - 3$ months and $3 - 9$ months which were converted into costs (Romeo et al., 2013). Details about the objective and conclusions from the trial as well as the study CEA are reported elsewhere (Banerjee et al., 2013). We restrict our attention to the 9 months cost-utility analysis of placebo ($n_1 = 111$) versus mirtazapine ($n_2 = 108$) from the health and social care cost perspective.

Table 1 shows the missing data patterns of the utility and cost data ($U_j$, $C_j$) at baseline ($j = 1$), 3 months ($j = 2$) and 9 months ($J = 3$) follow-up.

*Table 1. Missingness patterns for the utility and cost variables in the SADD study. For each pattern and treatment group the corresponding number and proportions of participants are reported. $U_j$ and $C_j$ indicate utility or cost variable at time j, while* **-** *and* **X** *indicate an observed or missing variable, respectively.*

| Missing data patterns | Placebo ($N_1$=111) | Mirtazapine ($N_2$=108) | Total (N=219) |
|---|---|---|---|
| $U_1$ $U_2$ $U_3$ $C_1$ $C_2$ $C_3$ | $N_1$(%) | $N_2$(%) | N(%) |
| - - - - - - | 54(48%) | 47(43%) | 101(46%) |
| - X X - X X | 12(11%) | 17(16%) | 29(13%) |
| - X X - - - | 15(13%) | 14(13%) | 29(13%) |
| - - X - - - | 11(10%) | 13(12%) | 24(11%) |
| - - X - - X | 9(8%) | 8(7%) | 17(8%) |
| - X - - - - | 4(4%) | 4(4%) | 8(4%) |
| - X X - - X | 2(2%) | 2(2%) | 4(2%) |
| - - - - X X | 2(2%) | 3(3%) | 5(2%) |
| - - - - - X | 2(2%) | 0(0%) | 2(1%) |

About half of the trial participants had complete cost-effectiveness data, with utility values which tended to be more frequently missing than costs.

### 3.2 Cost-effectiveness analysis of SADD

We applied the LMM model described above to the SADD data to estimate the difference in QALY and cost, adjusting for baseline values (utility or cost). A set of 10,000 bootstrap replications were performed to derive cost-effectiveness results. All models were fitted in R, using the packages and functions described in Appendix A.1. The R and STATA code for implementing the methods are given in Appendix A.2 and Appendix A.3, respectively.

# 4 RESULTS

## 4.1 Descriptive statistics

Table 2 reports the empirical means and standard deviations associated with the utility and cost variables in each treatment group over the study period.

*Table 2. Observed means (and standard deviations) of the utility and cost variables at baseline, 3 months and 9 months follow-up for the placebo and mirtazapine group in the SADD trial.*

|  |  | **Placebo** | ($N_1$=111) | **Mirtazapine** | ($N_2$=108) |
|---|---|---|---|---|---|
| Utilities |  |  |  |  |  |
|  | Baseline | 0.670 (0.268) | N=111 | 0.688 (0.292) | N=108 |
|  | 3 months | 0.733 (0.286) | N=78 | 0.763 (0.267) | N=71 |
|  | 9 months | 0.734 (0.274) | N=62 | 0.827 (0.195) | N=64 |
| Costs |  |  |  |  |  |
|  | Baseline | 1514 (3153) | N=111 | 1191 (2256) | N=108 |
|  | 3 months | 1437 (3338) | N=97 | 1130 (1868) | N=88 |
|  | 9 months | 2146 (4401) | N=84 | 2550 (4288) | N=78 |

People in the mirtazapine group had higher mean utilities at all time points, lower mean costs at the first follow-up and higher mean costs at the last follow-up.

Table 3 reports the LMM-estimated marginal means of the utilities and costs at each follow-up by group, as well as QALYs and total costs. Incremental results, ICER and probability of cost-effectiveness (at an acceptance threshold of £25,000) between the two groups are also reported.

*Table 3. Linear mixed model results for the SADD trial. Point estimates and 95% confidence intervals for the marginal means and mean difference of the utilities and costs at 3 ($j = 2$) and 9 ($j = 3$) months follow-up as well as for the quality-adjusted life years (QALYs) and total costs computed over the duration of the study. Incremental results are also reported in terms of the incremental cost-effectiveness ratio (ICER) and cost-effectiveness probability at $k = £25,000$ per QALY gained. Costs are expressed in British pounds.*

|  | **Placebo** | | **Mirtazapine** | | **Incremental** | |
|---|---|---|---|---|---|---|
|  | estimate | 95% CI | estimate | 95% CI | estimate | 95% CI |
| $U_2$ | 0.731 | (0.675; 0.786) | 0.752 | (0.694; 0.810) | 0.021 | (-0.053; 0.096) |
| $U_3$ | 0.727 | (0.671; 0.783) | 0:781 | (0.721; 0.841) | 0.054 | (-0.024; 0.132) |
| QALYs | 0.540 | (0.510; 0.571) | 0:561 | (0.530; 0.593) | 0.021 | (-0.016; 0.059) |
| $C_2$ | 1,369 | (840; 1,899) | 1,252 | (695; 1,809) | -117 | (-866; 631) |
| $C_3$ | 2,092 | (1,226; 2,959) | 2,760 | (1,856; 3,664) | 668 | (-544; 1,880) |
| Total costs | 3,462 | (2,236; 4,688) | 4,012 | (2,735; 5,289) | 550 | (-1,156; 2,257) |

|  | Placebo | Mirtazapine | Incremental |
|---|---|---|---|
| ICER |  |  | 57,774 |
| Probability of cost-effectiveness |  |  | 34% |

The results suggest that the mirtazapine group is associated with higher utilities compared with the placebo group at all follow-ups, which leads to higher mean QALYs estimates. Although mean costs at 13 weeks are on average higher in the placebo group, the mirtazapine group is associated with relatively higher mean costs at 9 months as well as higher mean total costs. Overall, the mean difference in QALY was $0.021 (95\% CI: -0.016; 0.059)$, and difference in cost $£550 (95\% CI: -1,156; 2,257)$. We additionally compare the results in terms of mean QALYs and total costs obtained under LMM with those from two alternative analyses: CCA and MI, the latter producing estimates in line with those of LMM with standard errors reduced by about 13% for QALYs and 35% for total cost estimates compared to CCA (see Appendix B.2).

Figure 1 shows the associated CEP and CEAC. Mirtazapine appears to be slightly more effective and costly, but with important uncertainty, resulting in a probability of being cost-effective of around 50% at a threshold of £25,000 per QALY gained.

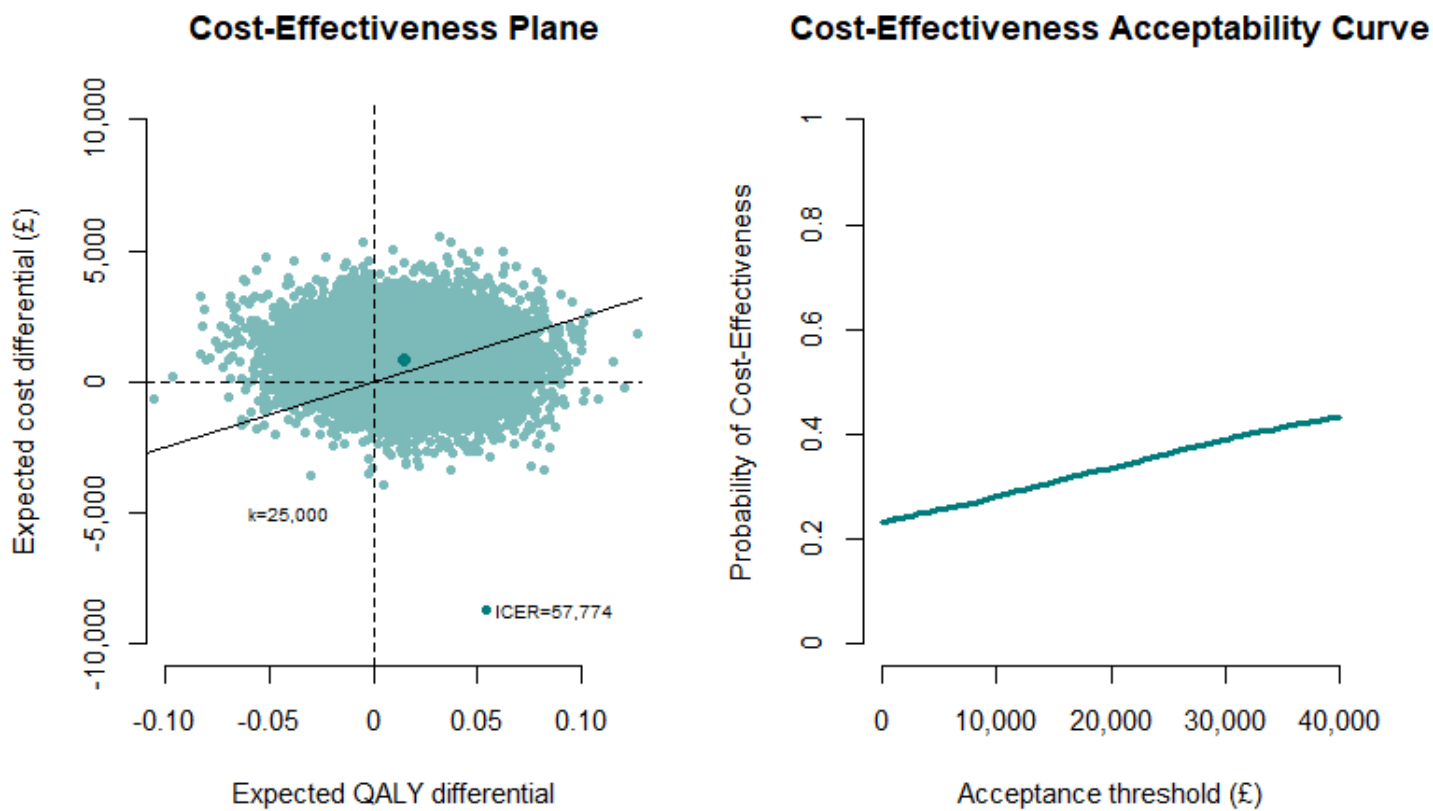

*Figure 1*. *CEP and CEAC from the SADD study, based on LMM fitted on 10,000 bootstrap replications. In the CEP, an acceptance threshold of $k =$ £25,000 per QALY gained is used and the position of the ICER on the plot is denoted with a darker coloured dot.*

## 5    DISCUSSION

In this article we proposed the use of LMMs as an alternative approach to conduct trial-based CEAs Although repeatedly criticised in the literature, analysts commonly handle missingness in QALYs and total costs via CCA (Gabrio et al., 2017; Leurent, Gomes, & Carpenter, 2018), which causes some loss of information and potentially bias the inferences. The use of imputation methods, especially multiple imputation, has been recommended in the literature since it can obtain valid inferences using all observed responses (Faria et al., 2014). However, there may be practical obstacles to the spread of these methods among practitioners. First, analysts may be unfamiliar with the implementation of multiple imputation, and feel more confident using simpler approaches. Second, multiple imputation can be time-consuming, particularly when combined with bootstrapping, with also alternative implementation strategies (Brand et al., 2019).

We believe LMMs represent an alternative approach which can overcome some of these limitations. First, practitioners may be more comfortable with the standard regression framework. Second, LMMs can be tailored to address other data features (e.g. cluster-randomised trials or non-normal distribution) while also easily combined with bootstrapping. Third, LMMs do not rely on imputation, and results are therefore deterministic and easily reproducible, whereas the Monte Carlo error associated with multiple imputation may cause results to vary from one imputation to another, unless the number of imputations is sufficiently large.

There are also limitations associated with the use of the proposed methods. First, LMM estimates are valid under MAR conditional on the observed outcome and the baseline variables in the analysis model. Multiple-imputation offers a more flexible framework allowing for inclusion of auxiliary variables not adjusted for in the analysis. Second, while MAR is often a reasonable starting point, sensitivity analysis under non-at-random assumptions should be considered (Gabrio et al., 2020; Leurent et al., 2020). While these can be conducted within a LMM framework (Daniels & Hogan, 2008), a multiple imputation or Bayesian framework is particularly well suited for this type of analyses (M. Gomes et al., 2019; Mason et al., 2018). Third, the LMM approach works well for repeatedly-measure outcomes, but addressing missingness within each component of the outcomes (e.g. for disaggregated cost components) increases the complexity of the model. Depending on the amount of observed data and time points, the suggested model may not achieve convergence, and simpler covariance structures could be considered.

Further work can be done to extend the proposed methods to tackle some of the typical features of CEA data. An interesting extension would be to fit a bivariate LMM model to model simultaneously the costs and utilities. Furthermore, while LMMs are robust to non-normally distributed data in large sample, generalised linear model specifications could be applied to improve the fit to the data using non-normal distributions (Nixon & Thompson, 2005). A possible solution is to explore the use of multivariate generalised LMMs to account for the correlation and characteristics of the data while also providing a coherent modelling framework based on the same number of individuals for both outcomes.

To conclude, we have shown how LMMs can be used for the analysis of trial-based CEAs Although the methodology illustrated is already known, particularly in the area of statistical analyses, to our knowledge LMMs have rarely been applied to health economic data collected alongside randomised trials. We believe the proposed methods is preferable to a complete-case analysis when CEA data are incomplete, and that it can offer an interesting alternative to imputation methods.

# APPENDIX

## A Implementation

### A.1 Software detail

All models can be fitted in `R` or `STATA` using different types of pre-defined functions and commands. In `R`, we use the function `lme` from the package `nlme`, which is specifically designed for fitting LMMs and produces standard statistical outputs, such as fixed and random effect estimates and their standard errors. Estimates for all model parameters are derived based on observed utility and cost responses under MAR using maximum likelihood methods. The function `emmeans` from the package `emmeans` can then be used to calculate point estimates and uncertainty measures for linear combinations of the model parameters, including the treatment effects as well as the marginal mean QALYs and total costs by treatment group. In `STATA`, we use the `mixed` command to fit the models and then use the post-processing commands `lincom` and `margins` to derive linear combinations of parameter estimates with associated uncertainty measures.

### A.2 R code

`R` code for fitting the utility and cost LMMs and derive the estimates for the CEA target quantities based on the data from the SADD study.

```
# load the R packages
library(nlme)
library(emmeans)
```

```r
# pre-process data

# load the dataset (in long format)
data_long_SADD <- readRDS("data_long_SADD.rds")
# variables: time (time indicator), trt (treatment indicator), u (utility values)
# 3 time points (baseline, 13 and 39 weeks follow-up)
# define treatment effect indicator at each follow-up
data_long_SADD$time_2<-ifelse(data_long_SADD$time==2,1,0)
data_long_SADD$time_3<-ifelse(data_long_SADD$time==3,1,0)
data_long_SADD$trt_time_2<-data_long_SADD$trt*data_long_SADD$time_2
data_long_SADD$trt_time_3<-data_long_SADD$trt*data_long_SADD$time_3
# define time as a factor
data_long_SADD$time <- factor(data_long_SADD$time)

# LMM for utilities

# fit the model using the lme function and use unstructured covariance
LMM_u<-lme(u ~ -1 + time + trt_time_2 + trt_time_3, random = ~ 1 | id,
                   data=data_long_SADD, method = "ML",
                   correlation = corSymm(form =~ as.numeric(time)|id),
                   weights=varIdent(form=~1|as.numeric(time)),
                   na.action = na.omit)
# look at parameter estimates and get confidence intervals
# fixed effects of interactions are the treatment effects at follow-ups
summary(LMM_u)
intervals(LMM_u, level = 0.95, which = "fixed")
# derive marginal means for the utilities in each group at each time point using the emme
ans function
mu_u <- emmeans(LMM_u, ~ -1 + time + trt_time_2 + trt_time_3)
# get confidence intervals
confint(mu_u)
# derive the marginal means for the QALYs in each group using the contrast function
# assign time weights to the marginals utility means to obtain correct mean QALYs
# (0.125 for baseline, 0.375 for 13 weeks follow-up, 0.25 for 39 weeks follow-up)
mu_e <- contrast(mu_u, list(mu0 = c(13/104,13/104 + 26/104,26/104,0,0,0,0,0,0,0,0,0),
                            mu1 = c(13/104,0,0,0,13/104 + 26/104,0,0,0,26/104,0,0,0)))
# derive mean differences between groups for the QALYs in a similar way
delta_e <- contrast(mu_e, list(diff = c(0,-13/104 - 26/104,-26/104,0,13/104 + 26/104,0,0,
0,26/104,0,0,0)))
# get confidence intervals
confint(mu_e)
confint(delta_e)

#LMM for costs

# fit the model using the lme function and use unstructured covariance
LMM_c <- lme(c ~ -1 + time + trt_time_2 + trt_time_3, random = ~ 1 | id,
              data=data_long_SADD, method = "ML",
              correlation = corSymm(form =~ as.numeric(time)|id),
              weights=varIdent(form=~1|as.numeric(time)),
              na.action = na.omit)
# look at parameter estimates and get confidence intervals
# fixed effects of interactions are the treatment effects at follow-ups
summary(LMM_c)
intervals(LMM_c, level = 0.95, which = "fixed")
```

```
# derive marginal means for the costs in each group at each time point using the emmeans
function
mu_c <- emmeans(LMM_c, ~ -1 + time + trt_time_2 + trt_time_3)
# get confidence intervals
confint(mu_c)
# derive the marginal means for the total costs in each group using the contrast function
mu_tc <- contrast(mu_c, list(mu0=c(0,1,1,0,0,0,0,0,0,0,0,0),
                             mu1=c(0,0,0,0,1,0,0,0,1,0,0,0)))
# derive mean differences between groups for the total costs in a similar way
delta_tc <- contrast(mu_tc, list(diff = c(0,-1,-1,0,1,0,0,0,1,0,0,0)))
# get confidence intervals
confint(mu_tc)
confint(delta_tc)
```

### A.3 STATA code

STATA code for fitting the utility and cost LMMs and derive the estimates for the CEA target quantities based on the data from the SADD study.

```
** Open data
        use "data_long_SADD.dta", clear
        gen trtp = trt*(time>1) //Treatment arm indicator, for follow-up visits only

** LMM for utilities
        mixed u i.time i.time#i.trtp || id:, res(unstructured, t(time))
                // || id: = random effect for each participant
                // res(unstructured, t(time)) = unstructured covariance, for each timepoint
                // i.time#i.trtp = interactions between time and treatment

        *Incremental utility at each time point:
                lincom 2.time#1.trtp   //Treatment effect (trtp=1) at time 2
                lincom 3.time#1.trtp   //Treatment effect at time 3
        *Incremental QALYs
                lincom 0.375*2.time#1.trtp + 0.25*3.time#1.trtp

        *Utility per arm at each time-point
                margins i.time#i.trtp
        *QALYs per arm
                margins i.time#i.trtp, post
                lincom 0.125*1.time#0.trtp + 0.375*2.time#0.trtp + 0.25*3.time#0.trtp
                lincom 0.125*1.time#0.trtp + 0.375*2.time#1.trtp + 0.25*3.time#1.trtp

** LMM for costs
                mixed c i.time i.time#trtp || id:, res(unstructured, t(time))

                *Incremental cost at each time point:
                        lincom 2.time#1.trtp   //Treatment effect at time 2
                        lincom 3.time#1.trtp   //Treatment effect at time 3
                *Incremental total cost
                        lincom 2.time#1.trt + 3.time#1.trt

                *Cost per arm at each time-point
                        margins i.time#trtp
                *Total cost per arm
```

```
                    margins i.time#trtp, post
                    lincom 2.time#0.trt + 3.time#0.trt
                    lincom 2.time#1.trt + 3.time#1.trt
```

# B Results
## B.1 Model estimates from the application to the SADD study

*Table 4*. Point estimates and 95% confidence intervals for each parameter from the utility and cost LMMs fitted to the data of the SADD study under MAR. The names of the independent variables associated with each parameter are reported.

|  | utilities model | | costs model | |
|---|---|---|---|---|
|  | estimate | 95% CI | estimate | 95% CI |
| **fixed effects** | | | | |
| $TIME_1$ | 0.674 | (0.637; 0.711) | 1355 | (991; 1720) |
| $TIME_2$ | 0.731 | (0.675; 0.786) | 1369 | (840; 1899) |
| $TIME_3$ | 0.727 | (0.671; 0.783) | 2092 | (1226; 2959) |
| $TIME_2$ TRT | 0.021 | (-0.053; 0.095) | -117 | (-866; 631) |
| $TIME_3$ TRT | 0.054 | (-0.024; 0.132) | 668 | (-544; 1880) |

## B.2 Parameters estimates under LMM, CCA and MI, fitted to the SADD data

*Table 5.* Estimates and standard errors (SE) for the QALYs and total cost marginal and incremental means under MAR for LMM, CCA and MI.

|  | Placebo | | Mirtazapine | | Incremental | |
|---|---|---|---|---|---|---|
|  | Estimate | SE | Estimate | SE | Estimate | SE |
| **CCA** | | | | | | |
| QALYs | 0.557 | (0.016) | 0.579 | (0.017) | 0.022 | (0.023) |
| Total costs | 3,375 | (774) | 3,695 | (831) | 320 | (1,148) |
| **MI** | | | | | | |
| QALYs | 0.541 | (0.013) | 0.561 | (0.014) | 0.021 | (0.020) |
| Total costs | 3,472 | (593) | 4,023 | (623) | 552 | (862) |
| **LMM** | | | | | | |
| QALYs | 0.540 | (0.016) | 0.562 | (0.016) | 0.021 | (0.019) |
| Total costs | 3,462 | (623) | 4,012 | (649) | 550 | (868) |

Parameter estimates under CCA and MI were retrieved using the following model specifications:

**CCA**. The analysis used only individuals with fully-observe utility and cost data at each time point in the SADD study, i.e. the 101 completers as shown in the first row of Table 1. The analysis was conducted using separate standard linear regression models fitted at the level of QALYs and total costs, including treatment and baseline outcome values as predictors to derive adjusted estimates for the incremental mean QALYs and total costs between the groups of the study.

**MI**. The analysis used all available observed utility and cost data at each time point in the SADD study and consisted in two steps. In the first step, imputation for missing utility and cost values was carried out using a multiple imputation by chained equation approach (Van Buuren, 2018). We imputed both outcomes under the assumption of normality to ensure a comparable match with the assumptions of LMM under MAR. The imputation models included observed outcome data at any time point as predictors, and imputation was performed by treatment group with a total number of 500 imputed datasets. Following current recommendations for trial-based CEA (Faria et al., 2014), we also consider imputation based on predictive mean matching which however lead to similar results. In the second step, within each imputed dataset, outcome variables were aggregated into QALY and total cost measures and standard linear regression models, as defined in CCA, were fitted to the imputed data and parameter estimates were derived. Finally, estimates were combined across the imputed datasets via Rubin's rules to correctly quantity missingness uncertainty.

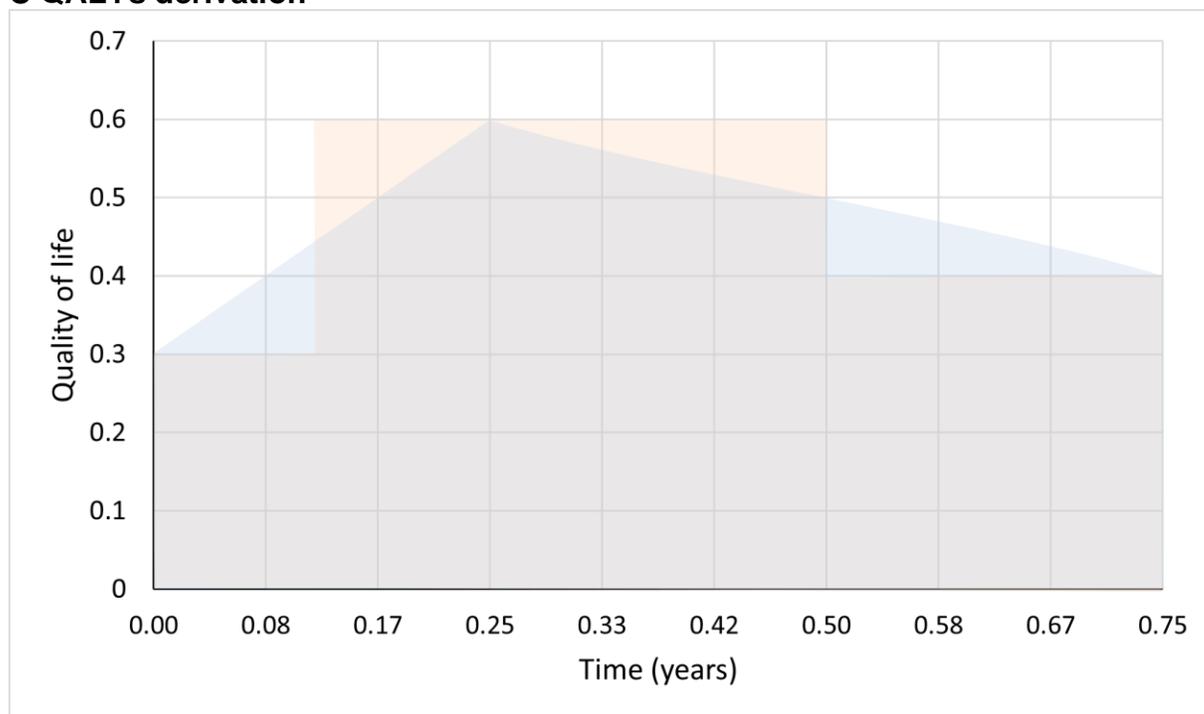

*Figure 2*. Derivation of QALYs as a weighted average of utilities

Figure 2 illustrates how to derive the QALYs defined by the area under the curve (AUC) of the utilities, for example for an individual with utilities of 0.3, 0.6 and 0.4 at baseline, 3 and 9 months respectively. Assuming a linear increase between utilities, the AUC is traditionally defined by the sum of multiple trapezoids aereas (in blue here). This can be rewriten as a sum of rectangular aereas (in pink here), defined by $t_{0.1} * u_0 + (t_{0.1} + t_{1.2}) * u_1 + (t_{1.2}) * u_2$, where $u_1$ is the utility at time $j$, and $t_{(i.j)}$ is half of the time (in years) between time-points $i$ and $j$. For example for this individual, $AUC = 0.125 * 0.3 + 0.375 * 0.6 + 0.25 * 0.4 = 0.3625$.

The same weights can be used to derive the mean AUC for a group of individuals (based on the group mean utility at each time-point), or the mean difference in AUC between two groups (based on the mean difference in utility at each time point). Discounting can be applied by reducing the relevant weights, for example after one year. Note that when deriving the difference in QALYS adjusted for baseline, there is no need to include the baseline term which is null.

# REFERENCES

Banerjee, S., Hellier, J., Dewey, M., Romeo, R., Ballard, C., Baldwin, R., Bentham, P., Fox, C., Holmes, C., Katona, C.others. (2011). Sertraline or mirtazapine for depression in dementia (HTA-SADD): A randomised, multicentre, double-blind, placebo-controlled trial. *The Lancet*, *378*(9789), 403–411.

Banerjee, S., Hellier, J., Romer, R., Dewey, M., Knapp, M., Ballard, C., Baldwin, R., Bentham, P., Fox, C., Holmes, C.others. (2013). Study of the use of antidepressants for depression in dementia: The HTA-SADD trial-a multicentre, randomised, double-blind, placebo-controlled trial of the clinical effectiveness and cost-effectiveness of sertraline and mirtazapine. *Health Technology Assessment*, *17*(7).

Black, W. C. (1990). The CE plane: A graphic representation of cost-effectiveness. *Medical Decision Making*, *10*(3), 212–214.

Brand, J., Buuren, S. van, Cessie, S. le, & Hout, W. van den. (2019). Combining multiple imputation and bootstrap in the analysis of cost-effectiveness trial data. *Statistics in Medicine*, *38*(2), 210–220.

Carpenter, J., & Kenward, M. (2012). *Multiple imputation and its application*. John Wiley & Sons.

Daniels, MJ., & Hogan, JW. (2008). *Missing data in longitudinal studies: Strategies for bayesian modeling and sensitivity analysis*. Chapman; Hall.

Dolan, P. (1997). Modeling valuations for EuroQol health states. *Medical Care*, 1095–1108.

Drummond, MF., Schulpher, MJ., Claxton, K., Stoddart, GL., & Torrance, GW. (2005). *Methods for the economic evaluation of health care programmes. 3rd ed*. Oxford university press.

Faria, A., Gomes, M., Epstein, D., & White, I. (2014). A guide to handling missing data in cost-effectiveness analysis conducted within randomised controlled trials. *PharmacoEconomics*, *32*, 1157–1170.

Gabrio, A., Daniels, M. J., Baio, G.others. (2020). A bayesian parametric approach to handle missing longitudinal outcome data in trial-based health economic evaluations. *Journal of the Royal Statistical Society Series A*, *183*(2), 607–629.

Gabrio, A., Mason, A. J., & Baio, G. (2017). Handling missing data in within-trial cost-effectiveness analysis: A review with future recommendations. *PharmacoEconomics-Open*, *1*(2), 79–97.

Gabrio, A., Mason, A., & Baio, G. (2019). A full Bayesian model to handle structural ones and missingness in economic evaluations from individual-level data. *Statistics in Medicine*, *38*, 1399–1420.

Gomes, M., Radice, R., Camarena Brenes, J., & Marra, G. (2019). Copula selection models for non-gaussian outcomes that are missing not at random. *Statistics in Medicine*, *38*(3), 480–496.

Gomes, R., Grieve, R., Nixon, R., & Edmunds, WJ. (2012). Statistical methods for cost-effectiveness analyses that use data from cluster randomized trials. *Medical Decision Making*, *32*, 209–220.

Leurent, B., Gomes, M., & Carpenter, J. R. (2018). Missing data in trial-based cost-effectiveness analysis: An incomplete journey. *Health Economics*, *27*(6), 1024–1040.

Leurent, B., Gomes, M., Cro, S., Wiles, N., & Carpenter, J. R. (2020). Reference-based multiple imputation for missing data sensitivity analyses in trial-based cost-effectiveness analysis. *Health Economics*, *29*(2), 171–184.

Leurent, B., Gomes, M., Faria, R., Morris, S., Grieve, R., & Carpenter, J. (2018). Sensitivity analysis for not-at-random missing data in trial-based cost-effectiveness analysis: A tutorial. *PharmacoEconomics*, 1–13.

Little, R. J., D'Agostino, R., Cohen, M. L., Dickersin, K., Emerson, S. S., Farrar, J. T., Frangakis, C., Hogan, J. W., Molenberghs, G., Murphy, S. A.others. (2012). The prevention and treatment of missing data in clinical trials. *New England Journal of Medicine*, *367*(14), 1355–1360.

Manca, A., Rice, N., Sculpher, M. J., & Briggs, A. H. (2005). Assessing generalisability by location in trial-based cost-effectiveness analysis: The use of multilevel models. *Health Economics*, *14*(5), 471–485.

Mason, A. J., Gomes, M., Grieve, R., & Carpenter, J. R. (2018). A bayesian framework for health economic evaluation in studies with missing data. *Health Economics*, *27*(11), 1670–1683.

Molenberghs, G., Thijs, H., Jansen, I., Beunckens, C., Kenward, M. G., Mallinckrodt, C., & Carroll, R. J. (2004). Analyzing incomplete longitudinal clinical trial data. *Biostatistics*, *5*(3), 445–464.

Nixon, R., & Thompson, S. (2005). Methods for incorporating covariate adjustment, subgroup analysis and between-centre differences into cost-effectiveness evaluations. *Health Economics*, *14*, 1217–1229.

Noble, S., Hollingworth, W., & Tilling, K. (2012). Missing data in trial-based cost-effectiveness analysis: The current state of play. *Health Economics*, *21*, 187–200.

Ramsey, S. D., Willke, R. J., Glick, H., Reed, S. D., Augustovski, F., Jonsson, B., Briggs, A., & Sullivan, S. D. (2015). Cost-effectiveness analysis alongside clinical trials II—an ISPOR good research practices task force report. *Value in Health*, *18*(2), 161–172.

Rice, N., & Jones, A. (1997). Multilevel models and health economics. *Health Economics*, *6*(6), 561–575.

Romeo, R., Knapp, M., Hellier, J., Dewey, M., Ballard, C., Baldwin, R., Bentham, P., Burns, A., Fox, C., Holmes, C.others. (2013). Cost-effectiveness analyses for mirtazapine and sertraline in dementia: Randomised controlled trial. *The British Journal of Psychiatry*, *202*(2), 121–128.

Rubin, DB. (1987). *Multiple imputation for nonresponse in surveys*. John Wiley; Sons.

Schafer, J. L., & Graham, J. W. (2002). Missing data: Our view of the state of the art. *Psychological Methods*, *7*(2), 147.

Van Buuren, S. (2018). *Flexible imputation of missing data*. Chapman; Hall/CRC.

Van Hout, B. A., Al, M. J., Gordon, G. S., & Rutten, F. F. (1994). Costs, effects and c/e-ratios alongside a clinical trial. *Health Economics*, *3*(5), 309–319.

White, I. R., & Thompson, S. G. (2005). Adjusting for partially missing baseline measurements in randomized trials. *Statistics in Medicine*, *24*(7), 993–1007.

Wolfinger, R. (1993). Covariance structure selection in general mixed models. *Communications in Statistics-Simulation and Computation*, *22*(4), 1079–1106.